\documentclass[aps,prc,reprint,superscriptaddress,showpacs]{revtex4-1}
\usepackage{graphicx}% Include figure files
\usepackage{dcolumn}% Align table columns on decimal point
\begin{document}

%\preprint{}

\title{Extinction of the $N=20$ neutron-shell closure for $^{32}$Mg examined by direct mass measurements
%\\Lowest strength of nuclear shell closure for magic nuclides
}


\author{A. Chaudhuri}
\email[]{ankur@triumf.ca}
%\thanks{}
\affiliation{TRIUMF, 4004 Wesbrook Mall, Vancouver, BC V6T 2A3, Canada}
\author{C. Andreoiu}
\affiliation{Department of Chemistry, Simon Fraser University, Burnaby, BC V5A 1S6, Canada}
\author{T. Brunner}
\affiliation{TRIUMF, 4004 Wesbrook Mall, Vancouver, BC V6T 2A3, Canada}
\affiliation{Department of Physics, Stanford University, Stanford, CA 94305, USA}
\author{U. Chowdhury}
\affiliation{TRIUMF, 4004 Wesbrook Mall, Vancouver, BC V6T 2A3, Canada}
\affiliation{Department of Physics and Astronomy, University of Manitoba, Winnipeg, MB R3T 2N2, Canada}
\author{S. Ettenauer}
\altaffiliation[Present address: ]{Department of Physics, Harvard University, Cambridge, MA 02138, USA}
\affiliation{TRIUMF, 4004 Wesbrook Mall, Vancouver, BC V6T 2A3, Canada}
\affiliation{Department of Physics and Astronomy, University of British Columbia, Vancouver, BC V6T 1Z1, Canada}
\author{A.T. Gallant}
\affiliation{TRIUMF, 4004 Wesbrook Mall, Vancouver, BC V6T 2A3, Canada}
\affiliation{Department of Physics and Astronomy, University of British Columbia, Vancouver, BC V6T 1Z1, Canada}
\author{G. Gwinner}
\affiliation{Department of Physics and Astronomy, University of Manitoba, Winnipeg, MB R3T 2N2, Canada}
\author{A.A. Kwiatkowski}
\affiliation{TRIUMF, 4004 Wesbrook Mall, Vancouver, BC V6T 2A3, Canada}
\author{A. Lennarz}
\affiliation{TRIUMF, 4004 Wesbrook Mall, Vancouver, BC V6T 2A3, Canada}
\affiliation{Institut f{\"u}r Kernphysik, Westf{\"a}lische Wilhelms-Universit{\"a}t, 48149 M{\"u}nster, Germany}
\author{D. Lunney}
\affiliation{CSNSM-IN2P3-CNRS, Universit\'{e} de Paris Sud, 91405 Orsay, France}
\author{T. D. Macdonald}
\affiliation{TRIUMF, 4004 Wesbrook Mall, Vancouver, BC V6T 2A3, Canada}
\affiliation{Department of Physics and Astronomy, University of British Columbia, Vancouver, BC V6T 1Z1, Canada}
\author{B.E. Schultz}
\affiliation{TRIUMF, 4004 Wesbrook Mall, Vancouver, BC V6T 2A3, Canada}
\author{M.C. Simon}
\affiliation{TRIUMF, 4004 Wesbrook Mall, Vancouver, BC V6T 2A3, Canada}
\author{V.V. Simon}
\altaffiliation[Present address:]{Helmholtz-Institut Mainz, 55128 Mainz, Germany}
\affiliation{TRIUMF, 4004 Wesbrook Mall, Vancouver, BC V6T 2A3, Canada}
\affiliation{Max-Planck-Institut f{\"u}r Kernphysik, 69117 Heidelberg, Germany}
%\affiliation{Fakul{\"a}t f{\"u}r Physik und Astronomie, Ruprecht-Karls-Universit{\"a}t Heidelberg, 69120 Heidelberg, Germany}
\author{J. Dilling}
\affiliation{TRIUMF, 4004 Wesbrook Mall, Vancouver, BC V6T 2A3, Canada}
\affiliation{Department of Physics and Astronomy, University of British Columbia, Vancouver, BC V6T 1Z1, Canada}

\date{\today}

\begin{abstract}
The `island of inversion' around $^{32}$Mg is one of the most important paradigm for studying the disappearance of the stabilizing `magic' of a shell closure. We present the first Penning-trap mass measurements of the exotic nuclides $^{29-31}$Na and $^{30-34}$Mg, which allow a precise determination of the empirical shell gap for $^{32}$Mg. The new value of 1.10(3) MeV is the lowest observed shell gap for any nuclide with a canonical magic number.
\end{abstract}
%\keywords{}

\maketitle
%
%\section{Introduction}
%
The building blocks of our knowledge of the atomic nucleus are the liquid drop model and the nuclear shell model \cite{1}. The former characterizes incompressible nuclear matter with short-range saturation of the nuclear force while the latter describes a sequence of quantum-mechanical single-particle orbital (labeled $s,p,d,f,$ etc.). The most salient feature of the shell model is the existence of magic numbers of neutrons or protons ($N$ or $Z=2, 8, 20, 28, 50, 82,$ and $N=126$) corresponding to closed shells that form nuclides of exceptional stability and spherical shape. The strength of a nuclear shell closure is indicated by the energy gap between the closed shell and the next available orbital. Exceptionally strong shell closures laid the foundation for the nuclear shell model discovered with stable nuclides, but these magic numbers have shown an elusive character with radioactive nuclides. For these unstable nuclides, known magic numbers can vanish, and new ones have appeared \cite{2,3}.
\\\\
The case of $N=20$ is a crucial benchmark for the disappearance of the stabilizing effects of closed shells. Anomalies in the nuclear structure in the region around the neutron magic number $N=20$ for $A \approx 32$ have been experimentally known for many years \cite{4,5}. The first indications of the breakdown of the $N=20$ shell closure were revealed from the on-line mass spectrometry studies of Na ($Z=11$) across the $N=20$ shell closure  \cite{4}. The masses of $^{31}$Na and $^{32}$Na obtained from \cite{4} implied that $N=20$ isotones neighboring $^{32}$Mg are more tightly bound than expected from the standard shell model predictions. The breakdown of the $N=20$ shell closure in this region originates in nuclear deformation and the resultant inversion of the standard $sd$-shell configuration and $pf$-shell intruder configuration, which has led to this region in the nuclear chart being known as the `island of inversion' \cite{6}. Here the intruder states imply the states comprised of shell-model configurations with
 $1p1h$, $2p2h$ or higher excited configurations across the $N=20$ shell gap \cite{7}.
\\\\
Immediately following the first conventional mass measurements, nuclear decay spectroscopy offered more evidence of an unexpected onset of deformation near $N=20$ \cite{5}. Coulomb excitation studies of $^{32}$Mg also confirmed the large deformation and pointed to the vanishing of the $N=20$ shell gap \cite{8}. Spectroscopic measurements continued to provide important information about the extent of the island of inversion \cite{9,10}. Laser spectroscopy has revealed complementary details about the scale of the deformation, and the size and limit of the island of inversion \cite{11}. Developments with radioactive beams have added to the wealth of experimental data and now allow precise probing of different levels by selective reactions, as illustrated by the discovery of a deformed $0^{+}$ state in $^{32}$Mg from a transfer reaction \cite{12}.\\\\
On the theoretical side, the change in the shell structure at $N=20$ is now largely attributed to the monopole effect of tensor forces \cite{13,14}. In this picture, the reduced attraction of the proton $d_{5/2}$ orbital towards the dripline allows the neutron $d_{3/2}$ orbital to increase in energy, reducing the gap to the $f_{7/2}$. As a consequence, a new gap arises with respect to the $s_{1/2}$ orbital for $N=16$, confirmed by the recent results showing the double magicity of $^{24}$O \cite{2,3}.
\\\\
Direct mass measurements can provide a model-independent value of nuclear binding energy, and hence a direct determination of the strength of the nuclear shell closure which is independent of any theoretical viewpoint. Penning traps confine charged particles to a small volume with well-defined electric and magnetic fields, making them the tool of choice for high-accuracy and -precision mass measurements of radioactive isotopes. For this reason, several Penning trap mass spectrometers have been installed at RIB facilities worldwide \cite{15}. In particular, the beam production, advanced ion manipulation, and mass measurement must be compatible with the decay lifetime of the radioactive ions, here $17-313$ ms.  With the TITAN experiment \cite{16}, Penning-trap mass spectrometry has developed to the point where precision mass measurements of exotic short-lived species ($t_{1/2} \sim 10-50$ ms) can now achieve a new level of precision and accuracy for the binding energy \cite{17}. In this article we report the first Penning-trap results on the island-of-inversion nuclides $^{29-31}$Na and $^{30-34}$Mg employing the TITAN spectrometer. 
\\\\
The TITAN Penning trap mass spectrometer is coupled to the rare-isotope-beam facility ISAC \cite{18} at TRIUMF. TITAN facility consists of three ion traps: a gas-filled radiofrequency quadrupole ion trap (RFQ) \cite{19} for cooling and bunching of ions, an electron beam ion trap (EBIT) \cite{20} for charge breeding \cite{21}, and a precision Penning trap for the mass measurements. The short-lived nuclides studied here were produced for the first time by impinging a 10-$\mu$A proton beam on a UC$_2$ target and employing a resonant laser ionization scheme \cite{22} at ISAC. The yield varied from $\approx 10^6$ particles per second for $^{30}$Mg to $\approx 10^3$ particles per second for more exotic $^{34}$Mg. 
\begin{figure}
\hspace*{0.1in}
\includegraphics[width=1.1\linewidth]{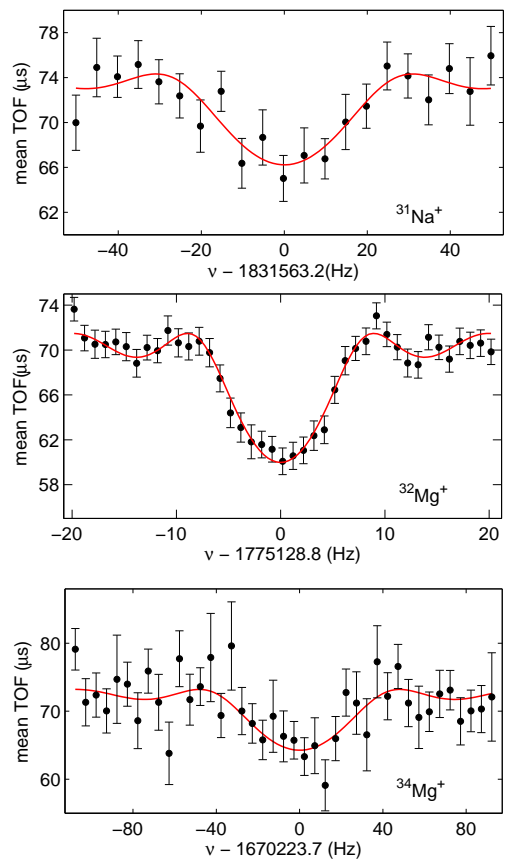}
\caption{\label{fig1} Time-of-flight ion-cyclotron-resonances for $^{31}$Na$^+$ ($t_{1/2}=17$ ms), $^{32}$Mg$^+$ ($t_{1/2}=86$ ms), and $^{34}$Mg$^+$ ($t_{1/2}=20$ ms) ions. The TOF is shown as a function of the excitation frequency for excitation times of $28$ ms, $97$ ms, and $18$ ms, respectively applied in the TITAN Penning trap. The solid line represents a fit of the theoretically expected resonance curve \cite{24} to the data points. The centroid corresponds to the cyclotron frequency, from which the mass is deduced.}
\end{figure}
\begin{figure}[bp]
\includegraphics[width=1.05\linewidth]{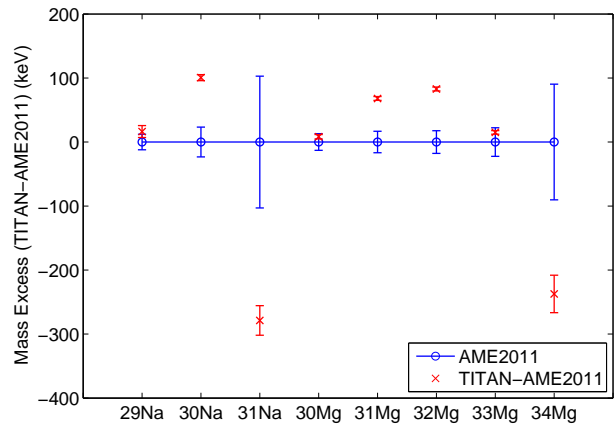}
\caption{\label{fig2} Comparison of the mass excess from the present work to the values from AME2011 \cite{33}. The present TITAN mass measurements are already included in the new AME2012 \cite{27}.}
\end{figure}
Radioactive ion beams delivered by ISAC at energy $\approx 20$ keV were injected into the RFQ cooler and buncher, where the ions were cooled, and then ejected as a low-emittance, bunched beam. Due to the short half-lives, the singly-charged ions extracted from the RFQ were sent to the Penning trap for the mass measurements, bypassing the EBIT. A Bradbury-Nielsen gate \cite{23} was used for selecting the ions of a specific mass-to-charge ratio prior to the injection into the measurement Penning trap.
\\\\
\begin{table*}
\caption{\label{table1}Summary of the results of the TITAN mass measurements of $^{29-31}$Na and $^{30-34}$Mg. The mass excess $ME$ with one-standard-deviation uncertainty is compared with the $ME$ from AME2003 \cite {30}, post-AME 2003 experiments \cite {31,32}, AME2011 \cite {33}, and AME2012 \cite {27}. Note that the given uncertainties in parentheses are the total uncertainties. The sources of the post-AME2003 experiments are cited in brackets. Note the AME2012 entries are dominated by the present TITAN measurements.}
\begin{ruledtabular}
\begin{tabular}{cccccccccc}
 Nuclide&$T_{1/2}$&Reference&$r$&$ME_{\rm TITAN}$&$ME_{\rm AME2003}$&$ME_{\rm expt}$&&$ME_ {\rm AME2011}$&$ME_{\rm AME2012}$\\
 &(ms)&&&(keV/c$^2$)&(keV/c$^2$)&(keV/c$^2$)&&(keV/c$^2$)&(keV/c$^2$)\\ \hline
 $^{29}$Na$^+$&$44.1$&$^{39}$K$^+$&$1.34344900(46)$ &$2686.0(9.3) $&$2665(13) $&$2669(12) $&$[31] $&$2670(12) $&$2680(7) $ \\
 $^{30}$Na$^+$&$48.4$&$^{39}$K$^+$ &$1.29840205(52) $&$8471.7(11.3)$&$$&$ $&$$&$$ \\
 $^{30}$Na$^+$&48.4&$^{16}$O$_2$$^+$ &$1.06600554(20)$&$8475.4(5.2)$&$ $&$ $&$$&$ $ \\
 $^{30}$Na$^+$&48.4&&&$8474.8(4.7)\footnote{weighted average of mass excesses obtained from two independent experiments using $^{39}$K$^+$ and $^{16}$O$_2$$^+$ as reference ions respectively }$&$8361(25)$&$8375(23) $&$[31]  $&$8374(23) $ &$8475(5)$\\
 $^{31}$Na$^+$&$17.0$&$^{39}$K$^+$ &$1.2563648(10)$&$12261(23)$&$12650(210) $&$12520(110) $&$ [32] $&$12540(103)$&$ 12261(23)$ \\
 $^{30}$Mg$^+$&$313$&$^{16}$O$_2$$^+$&1.06666796(13)&$-8883.8(3.4)$&$-8911(8) $&$-8892(13) $&$[31] $&$-8892(13)$&$-8884(3) $ \\
 $^{31}$Mg$^+$&$232$&$^{16}$O$_2$$^+$ &$1.03204213(11) $&$-3122.3(3.0)$&$-3217(12)$&$-3190(16) $&$[31]$&$ -3190(17)$&$-3122(3)$ \\
 $^{32}$Mg$^+$&$86$&$^{16}$O$_2$$^+$ &0.99970996(11)&$-828.8(3.1)$&$ -955(18)$&$-915(20) $&$ [31] $&$-912(18) $&$-829(3)$ \\
 $^{33}$Mg$^+$&$90.5$&$^{16}$O$_2$$^+$ &$0.96923179(9)$&$4962.2(2.8)$&$4894(20) $&$4947(22) $&$[31]  $&$4947(22)$&$4962.2(2.9) $ \\
 $^{34}$Mg$^+$&$20$&$^{16}$O$_2$$^+$ &$0.94062917(87)$&$8323(29)$&$8810(230) $&$8560(90) $&$ [32] $&$8560(90) $&$8323(29)$ \\
\end{tabular}
\end{ruledtabular}
\end{table*}
In Penning trap mass spectrometry, the atomic mass is determined from the measurement of the ion cyclotron frequency
\begin{equation}
\nu_{c}= \frac{1}{2\pi}\ \left(\frac{q}{m_{\rm ion}}\right) B,
\label{3}
\end{equation}
where $B$ is the magnetic field strength, $q$ is the charge, and $m_{\rm ion}$ is the mass of the ion. The cyclotron frequency of the trapped ion was measured by the time-of-flight ion-cyclotron-resonance (TOF-ICR) detection method \cite{24}. TOF resonances of $^{31}$Na$^+$, $^{32}$Mg$^+$, and $^{34}$Mg$^+$ are shown in Figure \ref{fig1}. To calibrate the magnetic field, $\nu_{c}$ of the ion of interest is compared to the $\nu_{c,{\rm ref}}$ of a reference ion with a well known mass $m_{\rm ref}$ i.e.
\begin{equation}
\frac{m_{\rm ref}}{m_{\rm ion}}\ = \left(\frac{\nu_{c}}{\nu_{c,\rm ref}}\right) \left(\frac{q_{\rm ref}}{q}\right) =r \left(\frac{q_{\rm ref}}{q}\right).
\label{4}
\end{equation}
Here $r=\frac{\nu_{c}}{\nu_{c,\rm ref}}$ is the cyclotron frequency ratio. A count-rate class analysis \cite{25} was carried out for all frequency measurements to account for possible systematic shifts in the cyclotron frequency and exclude the influences of any contaminant ion species in the Penning trap.
\\\\
Either $^{39}$K$^+$ ($\delta M \approx \rm$ 5 eV/c$^2$) ions obtained from a surface ion source coupled to TITAN, or $^{16}$O$_2^+$ ($\delta M \approx \rm$ 0.3 eV/c$^2$) ions obtained from the ISAC off-line ion source OLIS \cite{26} were used as reference ions during the measurements (where $\delta M$ is the uncertainty of the atomic mass taken from the 2012 atomic mass evaluation (AME) \cite{27}). The atomic mass $M$ of each short-lived nuclide was calculated from its measured frequency ratio $r$ and the reference ion's mass and was corrected for the electron mass. The electron binding energy of less than $10$ eV was neglected for the present measurements, whose statistical mass uncertainties are above $1$ keV. The measurement results are summarized in Table \ref{table1} in terms of the mass excess $=M-A$ in keV/c$^2$ where $M$ is the atomic mass, and $A$ is the atomic mass number in atomic mass units (u) taken from the 2010 CODATA-recommended values \cite{28}. Since the reference mass values can change in the future, the measured frequency ratio $r$ is listed in the Table \ref{table1}.
\begin{figure}
\includegraphics[width=1.05\linewidth]{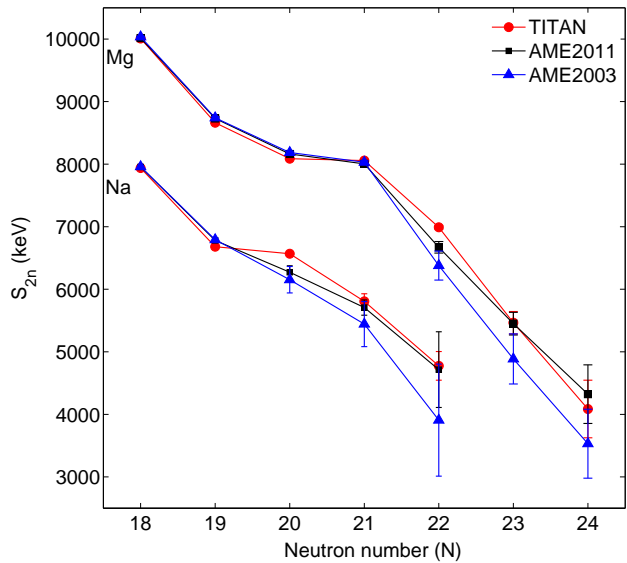}
\caption{\label{fig3} Two-neutron separation energy S$_{2n}$ plots of Mg, and Na isotopes. The solid circles represent the S$_{2n}$ values determined from the present work. The squares and the triangles denote S$_{2n}$ values from AME2011 \cite{33} and AME2003 \cite{30}, respectively.}
\end{figure}
\\\\
Large deviations were found in the mass excess values from past measurements as shown in Table \ref{table1} and illustrated in Figure \ref{fig2}. The time-of-flight measurements used in the past were prone to systematic errors due to the complex calibration and error evaluation technique. Often they used calibrants far from the stable region of the nuclear chart, hence re-evaluation was needed when the reference masses changed. The TITAN mass values also represent significant improvement in the resulting precision.
\\\\
The two-nucleon separation energy, an important quantity for probing the evolution of shell closures, is derived from the experimental mass data. The two-neutron separation energy
\begin{eqnarray}
S_{2n}(Z,N)=B(Z,N)-B(Z,N-2) \nonumber
\\
=-M(Z,N)+M(Z,N-2)+2m_n,
\end{eqnarray}
where $B(Z,N)$ is the binding energy, $M(Z,N)$ is the atomic mass and $m_n$ is the neutron mass, is a key measure of how tightly valence neutrons are bound to the nucleus and hence, is sensitive to the neutron-shell closure. The relative strength of the difference in binding energies before and after a purported shell closure can be quantified by an empirical shell gap, as defined by \cite{29}
\begin{equation}
\Delta_n(Z,N)=S_{2n}(Z,N)-S_{2n}(Z,N+2).
\label{2}
\end{equation}
The empirical shell gap $\Delta_n(Z,N)$ is a sensitive probe to the relative enhancement in binding energy found in magic nuclides, and is directly determined from mass measurements. Note that the empirical shell gap does not represent the energy difference between effective single-particle energies, a signature of shell evolution that is model-dependent. As a double derivative, $\Delta_n(Z,N)$ involves the masses of three nuclides as seen from Equation (3) and Equation (4), requiring high measurement precision to minimize its uncertainty. The masses of $^{30,32,34}$Mg and $^{29,31,33}$Na are needed for the evaluation of the of empirical shell gap for $^{32}$Mg and $^{31}$Na, respectively. The uncertainties associated with earlier conventional mass measurements \cite{31,32,33} were too large (see Figure \ref{fig2}) for a reliable determination of the empirical shell gap.
\\\\
\begin{figure}
%\vspace*{-.25in}
%\setlength{\abovecaptionskip}{-40pt}
%\setlength{\belowcaptionskip}{0pt}
\includegraphics[width=1\linewidth]{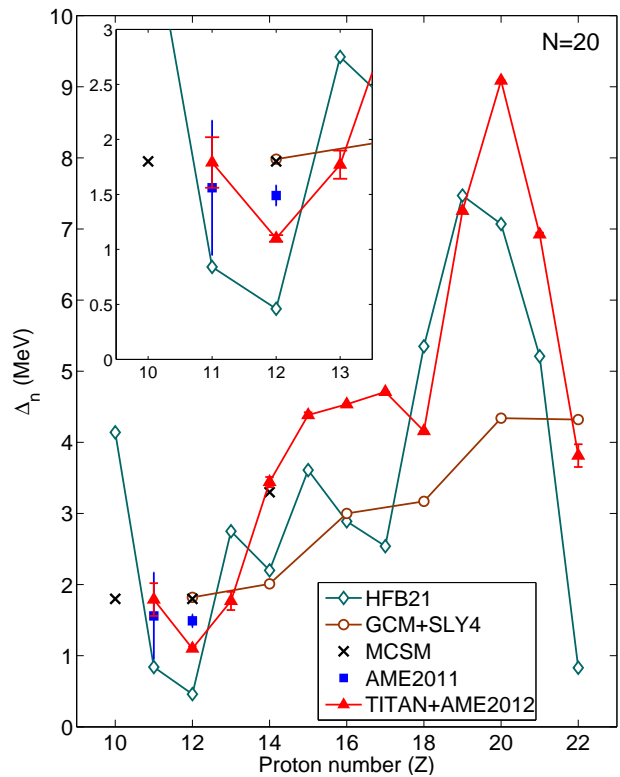}
\caption{\label{fig4} Comparison of experimental and theoretical empirical neutron shell-gap $\Delta_n$ for nuclides with magic neutron number $N=20$. The solid triangles represent the experimentally obtained values from TITAN and AME2012. The TITAN measurements contribute to the cases of $^{32}$Mg ($Z=12, N=20$) and $^{31}$Na ($Z=11, N=20$).  The $\Delta_n$ values obtained from HFB21, the angular-momentum projected generator method with the Skryme interaction SLy4 (GCM+SLy4), and the Monte Carlo Shell Model (MCSM) are shown in open diamonds, open circles, and crosses respectively. The $\Delta_n$ values of $^{32}$Mg and $^{31}$Na from AME2011 are shown in solid rectangles for comparison. $\Delta_n$ for $Z$=10 to 13 are shown in inset.}
\end{figure}
The new S$_{2n}$ values for Mg and Na isotopes are plotted in Figure \ref{fig3} and are compared with AME2003 \cite{30} and AME2011 \cite{33}. We compare to AME2011 rather than the new AME2012 \cite{27} because the present TITAN mass measurements are included in the AME2012 (see Table \ref{table1}). Note that in addition to the present TITAN mass values, the mass values of $^{28,29,35,36}$Mg, and $^{27,28,32}$Na from AME2012 and that of $^{33}$Na from the recent experiments \cite{32,34} were used to determine the new S$_{2n}$ values.
\\\\
The new mass values provide information relevant to the breakdown of the established nuclear shell model in the vicinity of $^{32}$Mg. The empirical shell gap $\Delta_n$ for $^{32}$Mg ($Z=12, N=20$) was determined using the atomic masses of $^{30,32,34}$Mg reported here. The value obtained is only $1.10(3)$ MeV, which reveals the lowest observed strength of nuclear shell closure for a nuclide with a conventional magic number. The present measurements of $^{29,31}$Na improve the precision of $\Delta_n$ for $^{31}$Na $(Z=11, N=20)$. The $\Delta_n$ for $^{31}$Na was calculated to be $1.79(23)$ MeV.
\\\\
Figure \ref{fig4} displays a comparison of the empirical shell gap $\Delta_n$ from $Z=10$ to $22$ for the magic number $N=20$. Enhanced stability demonstrated by an exceptionally high $\Delta_n$ value can be seen for doubly-magic $^{40}$Ca ($Z=N=20$) due to additional like-nucleon pairing. A significant drop of shell strength below $Z=14$ reveals the presence of the island of inversion. The empirical shell gap $\Delta_n$ using the mass values from AME2011 shows the trend of the $N=20$ shell closure; however, due to the large uncertainty it was impossible to come to any conclusion about the shell quenching. The first Penning trap measurements in this mass region using the TITAN experimental set-up confirm the vanishing of the $N=20$ shell closure purely from the direct atomic mass measurements.
\\\\
We compare our results (Figure \ref{fig4}) with the Hartree-Fock-Bogoliubov mass model (HFB21) using nonconventional Skyrme forces \cite{35}. The mean-field model HFB21 yields $\Delta_n$($^{32}$Mg)=$0.46$ MeV; however, this model predicts $^{32}$Mg to be spherical in contrast to the experimentally determined shape \cite{8,11}. We also compare our results with other theoretical studies: the angular-momentum projected generator-coordinate method with the Skryme interaction SLy4 (GCM+SLy4) \cite{36} which predicts $\Delta_n$($^{32}$Mg)=$1.82$ MeV, and the Monte Carlo Shell Model (MCSM) based on the quantum Monte Carlo diagonalization method \cite{37} which predicted  $\Delta_n$($^{32}$Mg)=$1.8$ MeV. The conventional shell model calculation \cite{38}, which is limited by the current knowledge of the interaction, predicted  $\Delta_n$($^{32}$Mg) to be negative. Despite the large differences in the $\Delta_n$ value, the standard shell model studies give a description of the anomaly which indicates the breaking of the shell closure at $N=20$. The $N=20$ region is challenging for energy-density functional methods \cite{39}, as it involves light nuclei, and also for {\em ab-initio} methods, which have not been able to access the island of inversion to date.
\\\\
The results from HFB21 and SLy4, while both based on the mean-field approach, complement each other. The parameters of the HFB21 interaction are fit to the entire mass table, resulting in a better overall fit to the shell gap values. The parameters of the SLy4 interaction are adjusted to a restricted set of (doubly magic) nuclides. Both models give similar extrapolations for the $N$=50 and $N$=82 shell closures and the tendency is the same for $N$=20. (Unfortunately there are no SLy4 predictions for $Z<$12.)
\\\\
There is discussion in literature (e.g. \cite{40}) considering the validity of the definition of the empirical shell gap given by Equation (4). In principle this definition should apply only to spherical nuclides, which we know $^{32}$Mg is not. Moreover, the shell gap defined this way reflects contributions from three different nuclides, spanning five neutron numbers. The definition using S$_{n}$ reduces the range to three neutron numbers but introduces the effect of pairing. But it is important to recall that $\Delta_n$ has the merit of being derived from experimental observables- unlike the effective single-particle energy, which is a theoretical quantity. Therefore the use of $\Delta_n$ does offer a comparative benchmark.
\\\\
The found 1.1-MeV shell gap is interesting in itself. The $N$=20 shell closure, so strong for $Z$=20, is extinguished for $Z$=12, reduced in strength by one order of magnitude. Given the higher value for $Z$=11, it is possible to think that the shell gap may continue its increase, rising for $Z$=10 through the (unbound) $Z$=8.
\\\\
Any theory correctly describing the magicity of the $N$=20 closure for $Z>$13 must also predict the binding energy differences for those $Z$ for which the shell is quenched.  While the mechanism of the tensor force provides an insightful qualitative picture, accurate models must be able to quantitatively predict the energies at which the competing structure effects act. Such quantitative ground-state predictions were provided by \cite{38,41} over ten years ago, but no new calculations have appeared since that time. The corresponding spectroscopic calculations also severely underestimated the 0$^+$ state discovered by \cite{12}. Moreover, the reaction codes used in that work required an adjustment to the binding energy of $^{32}$Mg. Recent success with three-body forces from chiral effective field theory \cite{42,43} show promise for accurate S$_{2n}$ predictions. We eagerly await such calculations for the island of inversion.
\begin{acknowledgments}
This work has been supported by the Natural Sciences and Engineering Research Council (NSERC) of Canada and the National Research Council (NRC) of Canada. A.T.G. acknowledges support from the NSERC CGS-D program, T.D.M. from the NSERC CGS-M program, A.L. from the Deutsche Forschungsgemeinschaft (DFG) under grant no. FR 601/3-1, S.E. from the Vanier CGS program, and V.V.S. from the Studienstiftung des deutschen Volkes. D.L. is supported by the French IN2P3. We thank TRIUMF operations team, in particular J. Lassen and the laser ion source group.
\end{acknowledgments}
{}
\end{document}